\documentclass[aps,prb,twocolumn,10pt,a4paper,showpacs]{revtex4-1}
\usepackage{graphicx}
\usepackage{amsmath}
\usepackage{amssymb}
\usepackage{bbm}

\begin{document}



\newcommand{\half}{\frac{1}{2}}
\newcommand{\ua}{\uparrow}
\newcommand{\da}{\downarrow}
\newcommand{\comment}[1]{}
\newcommand{\w}{\omega}
\newcommand{\pd}{\phantom\dagger}
\newcommand{\dmp}{\Delta_{\mathrm{mp}}(\w)}
\newcommand{\Xmpw}{X_{\mathrm{mp}}(\w)}
\newcommand{\Xtmpw}{\tilde{X}_{\mathrm{mp}}(\wtil)}
\newcommand{\Xmp}{X_{\mathrm{mp}}}
\newcommand{\dtmp}{\tilde{\Delta}_{\mathrm{mp}}(\tilde{\w})}
\newcommand{\etatil}{\tilde{\eta}}
\newcommand{\wtil}{\tilde{\w}}
\newcommand{\wtilp}{\tilde{\w}^{\prime}}
\newcommand{\y}{y(\tilde{\w})}
\newcommand{\dI}{\Delta_{I}^{\pd}(\w)}
\newcommand{\dJ}{\Delta_{J}^{\pd}(\w)}
\newcommand{\eI}{{\cal{E}}_{I}}
\newcommand{\eJ}{{\cal{E}}_{J}}
\newcommand{\ecal}{{\cal{E}}}

\newcommand{\n}[2][I]{n^{\scriptscriptstyle(#1)}_{#2}}
\newcommand{\Ntot}[2][I]{N^{\scriptscriptstyle(#1)}_{#2}}


\title{Simple probability distributions on a Fock-space lattice}


\author{Staszek Welsh and David E. Logan}
\affiliation{Department of Chemistry, Physical and Theoretical Chemistry, Oxford University, South Parks Road, Oxford, OX1 3QZ, United Kingdom}

\date{\today}

\begin{abstract}
We consider some aspects of a standard model employed in studies of many-body localization:
interacting spinless fermions with quenched disorder, for non-zero filling fraction, 
here on $d$-dimensional hypercubic lattices.  The model may be recast as an equivalent tight-binding model on a 
`Fock-space (FS) lattice' with an extensive local connectivity. In the thermodynamic limit exact 
results are obtained for the distributions of local FS coordination numbers, FS site-energies, and 
the density of many-body states. All such distributions are well captured by exact diagonalisation 
on the modest system sizes amenable to numerics. Care is however required in choosing the appropriate 
variance for the eigenvalue distribution, which has implications for reliable identification of mobility edges.
\end{abstract}


\maketitle


\section{Introduction}
\label{section:intro}

In recent years there has been great interest in the study of highly excited quantum states of disordered, 
interacting systems, and notably the phenomenon of many-body localization~\cite{BAAAnnPhys2006} (MBL);
for topical reviews see e.g.\ [\onlinecite{HuseReview2015},\onlinecite{AltmanVoskReview2015}].
In particular it is now well appreciated that in interacting systems, localization or its absence  
can be viewed quite generally as an Anderson localization problem in the Fock or Hilbert space of associated 
many-body states. This connection was made long ago in the context of the problem of quantum ergodicity in isolated
molecules,~\cite{DELPGWJCP1990,LeitnerReview2015} and later in quantum dots.~\cite{AltshuleretalPRL1997}

One of the central models~\cite{Oganesyan+HusePRB2007} extensively studied in MBL is that of interacting 
spinless fermions with quenched disorder, for non-zero fermion filling fractions. We consider it here, 
on $d$-dimensional hypercubic lattices. As  detailed in sec.\ \ref{section:model}, the model 
can be mapped exactly onto a tight-binding model on a `Fock-space lattice' with an extensive local 
connectivity, each site of which corresponds to a disordered, interacting many-body state with specified 
fermion occupancy of  all real-space sites, and is thus associated with a Fock-space `site-energy'; with 
Fock-space sites connected by the one-electron hopping matrix element of the original Hamiltonian.

While MBL \emph{per se} is not considered directly in this paper, we believe it is instructive to have some broad 
understanding of the Fock-space lattice  in a statistical sense. To that end we consider the distributions, over both 
Fock-space sites and (where relevant) disorder realisations, of the local Fock-space coordination numbers, site-energies, and many-body eigenvalues (or density of states). None of these quantities contain information about localization itself, but all reflect `basic' properties of the system in an obvious sense, and understanding them is a precursor to considering localization (to which we turn in a subsequent work~\cite{SWDEL2}). 

As expected from the central limit theorem all such distributions are normal, with extensive means; 
the relevant moments are obtained (sec.\ \ref{section:distributions}) as a function of the bare model parameters, fermion filling fraction and space dimension. A subtlety arising in the choice of variance for the eigenvalue spectrum (and Fock-space 
site-energies) is pointed out in sec.\ \ref{section:sec4}, with the correct choice enabling a pristine distinction between localized and extended states as a function of energy, as required for a reliable identification of mobility edges which may separate them. Comparison to exact diagonalisation shows the distributions considered to be well captured even for the small system sizes practically accessible to numerics in $d=1,2$.  Concluding remarks are given in sec.\ \ref{section:sec5}.


\section{Model}
\label{section:model}

The Hamiltonian is a standard model of spinless fermions,~\cite{Oganesyan+HusePRB2007}  
here considered on a $d$-dimensional lattice:
\begin{subequations}
\label{eq:1}
\begin{align}
H~=&~ H_{W}^{\pd}+H_{t}^{\pd}+H_{V}^{\pd}
\label{eq:1a}
\\
=& ~ \sum_{i} \epsilon_{i}^{\pd} \hat{n}_{i}^{\pd} +
\sum_{\langle ij \rangle} t ~(c_{i}^{\dagger}c_{j}^{\pd}+c_{j}^{\dagger}c_{i}^{\pd})
+\sum_{\langle ij \rangle} V~\hat{n}_{i}^{\pd}\hat{n}_{j}^{\pd}
\label{eq:1b}
\end{align}
\end{subequations}
with $\hat{n}_{i} =c_{i}^{\dagger}c_{i}^{\pd}$. The hoppings ($t$) and interactions ($V$) are nearest neighbour (NN), with 
$\langle ij \rangle$ denoting distinct NN pairs. The site energies $\{\epsilon_{i}\}$ in $H_{W}$
are characterised by quenched random disorder, with a distribution  $P(\epsilon_{i})$ common to all sites 
(and chosen for convenience to have zero mean, $\langle \epsilon \rangle =\int d\epsilon P(\epsilon)\epsilon  =0$).
The eigenvalues of $H$  are denoted by $E_{n}$.

The lattice has $N$ sites and contains $N_{e}$ fermions. We are interested in the thermodynamic limit where 
both $N\equiv L^{d}$ and $N_{e} \rightarrow \infty$, holding the filling $\nu = N_{e}/N$ fixed and non-vanishing. 
The coordination number of the lattice is denoted by $Z_{\mathrm{d}}$, with $Z_{\mathrm{d}}=2d$ for hypercubic lattices.


\subsection{Equivalent tight-binding model}
\label{subsection:TBM}

\begin{figure*}[t]
\includegraphics*{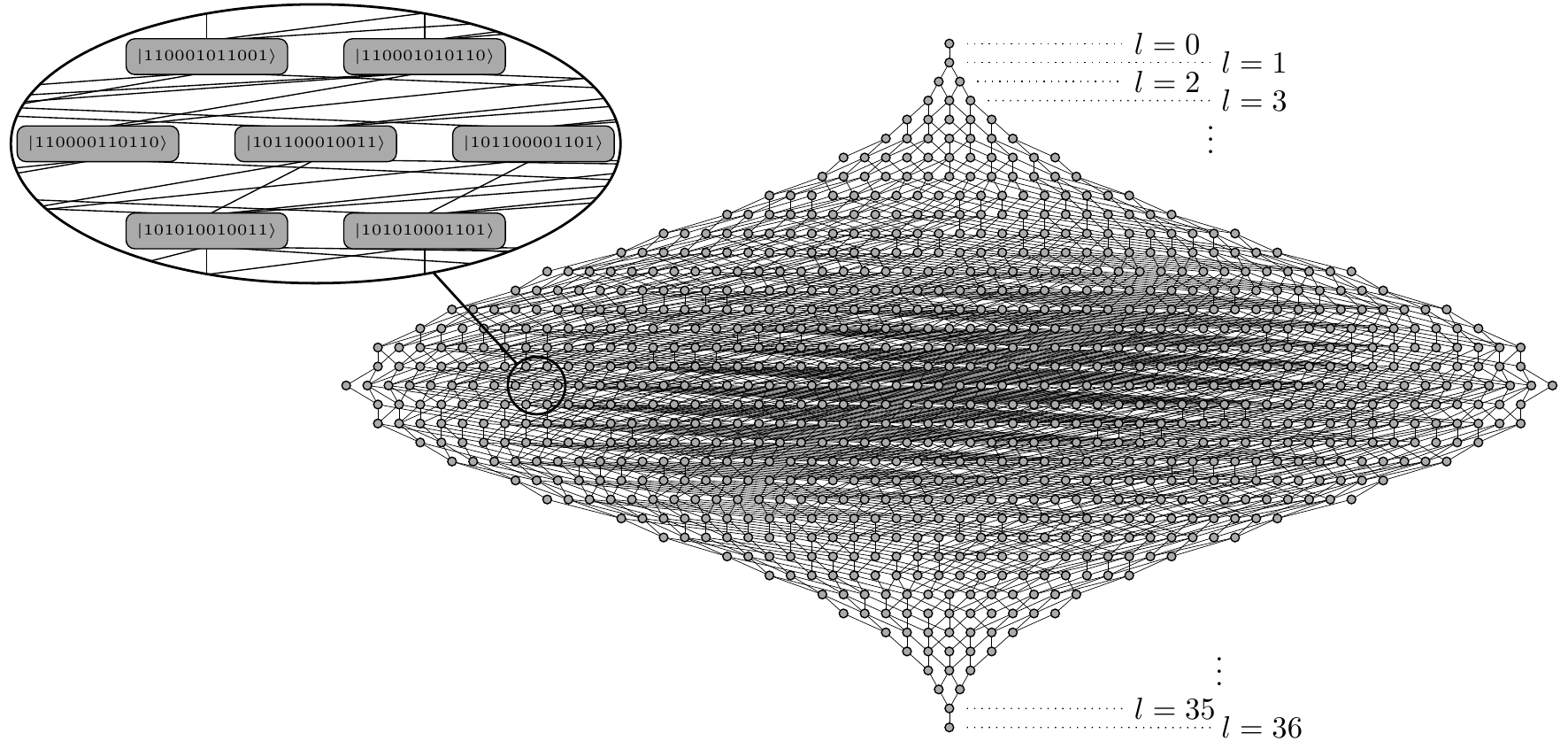}
\caption{\label{fig:fig1}
Example of Fock-space lattice (for an open $1d$ chain with $N=12$ sites and $N_{e}=6$ fermions, $N_{{\cal{H}}}=924$). Each site represents a state  $|I\rangle$, solid lines show coupling between states under hopping. See text for discussion. 
\emph{Inset}: a small section showing $7$ sites, with coordination numbers $Z_{I}$ ranging from $5-8$. 
}
\end{figure*}

For any given fermion number $N_{e}$, the dimension of the associated  Fock space (FS) is 
$N_{{\cal{H}}} = {^{N}}C_{N_{e}}$ ($\equiv \binom{N}{N_{e}}$), growing exponentially with the number of sites, 
$N_{{\cal{H}}} \propto N^{-1/2}e^{cN}$ (with $c= -[\nu\ln\nu +(1-\nu)\ln(1-\nu)]$ the configurational entropy per site).

The Hamiltonian may be recast as an effective tight-binding model (TBM) on a lattice of $N_{{\cal{H}}}$ `sites'.
To this end, first separate $H = H_{0}+H_{t}$ with $H_{0} = H_{W}+H_{V}$ (eq.\ \ref{eq:1}). Since $H_{0}$ involves solely number operators, its states $\{|I\rangle\}$ are
\begin{equation}
\label{eq:2}
|I\rangle =|\{\n{i}\}\rangle =(c_{1}^{\dagger})^{\n{1}}(c_{2}^{\dagger})^{\n{2}}....(c_{N}^{\dagger})^{\n{N}}|\mathrm{vac}\rangle
\end{equation}
with each occupation number $\n{i} =0$ or $1$ only, and $\sum_{i}\n{i} =N_{e}$ for any $|I\rangle$.
These states are orthonormal, with energies $\eI$ under $H_{0}$ given by 
\begin{equation}
\label{eq:3}
{\cal{E}}_{I}^{\pd} = ~\sum_{i} \epsilon_{i}^{\pd} \n{i} ~+~V r_{I}^{\pd},~~~~
r_{I}^{\pd}=~\sum_{\langle ij \rangle} ~\n{i}\n{j} 
\end{equation}
where $r_{I}$ is thus defined; and each of the $N_{{\cal{H}}}$ configurations of fermions on the lattice specifies 
uniquely one such basis state $|I\rangle$.

 The $\{|I\rangle\}$ may be viewed as forming a lattice  of $N_{{\cal{H}}}$ sites in state-space,~\cite{ DELPGWJCP1990}
here referred to as a `FS lattice'. These sites are connected under the hopping term $H_{t}$ (eq.\ \ref{eq:1}), with
$T_{IJ}=\langle I|H_{t}|J\rangle =T_{JI}$ the real symmetric matrix elements coupling them. Since the hopping  
is NN only, $|T_{IJ}| =t$ for $T_{IJ}$ non-vanishing ($T_{IJ}$ is either $\pm t$ for general $d\geq 2$, its
sign obviously depending on the configuration of fermions in $|I\rangle$ and $|J\rangle$; while $T_{IJ}=t$ for all 
connected FS sites in the $d=1$ open chain). The total Hamiltonian can thus be written as
\begin{equation}
\label{eq:4}
H~=~ \sum_{I} \eI^{\pd}|I\rangle\langle I|~+~
{\sum_{I,J (J\neq I)}}
T_{IJ}^{\pd}|I\rangle\langle J|  .
\end{equation}
This has the form of a TBM on the FS lattice; with the number of  sites to which any given site $I$ is connected under 
the hopping defining the  local coordination number of the site, denoted $Z_{I}$.

Given the mapping to an effective TBM, the same questions can obviously be asked about this FS lattice as arise for a 
one-electron disordered  TBM.~\cite{FN1} What for example is the disorder-averaged density of many-body states; and are those states of some given energy  localized on a vanishingly small fraction of the FS lattice, or delocalized over a finite fraction of it? The mapping also means that some techniques applicable to one-electron problems on a real-space 
lattice may be extended to encompass the question of MBL in the FS lattice (as we will consider in subsequent work~\cite{SWDEL2}).


\subsection{Fock space lattice}
\label{subsection:FSlattice}

We have found it helpful to have a concrete picture of the FS lattice, in part to prompt questions about its basic characteristics and  their implications.
An illustrative example is given in fig.\ \ref{fig:fig1}, shown for a small $1d$ open chain of $N=12$ sites, $N_{e}=6$ fermions (filling $\nu =1/2$).

Each circle represents a site (associated with a state $|I\rangle$), and solid lines denote connections between the states under the hopping. The sites are arranged in rows, with row index $l$. The top row ($l=0$) consists of the single state 
$|I\rangle =|L\rangle =|111111000000\rangle$ in which all fermions occupy the leftmost real-space sites. This state has a coordination number $Z_{I}=1$, since under $H_{t}$ it can connect only to the single state $|111110100000\rangle$ in row 
$l=1$.  The latter connects under $H_{t}$ to two further states in row $l=2$ (such that its coordination number $Z_{I}=3$); and these two states are connected in turn to a total of three states in row $l=3$. The process continues in this fashion, with the number of sites/states in successive layers first growing and then decreasing as seen in fig.\ \ref{fig:fig1}
(reflecting that a fermion cannot move beyond the end site for an open chain); and terminates at the bottom state
$|I\rangle =|R\rangle =|000000111111\rangle$ in which all fermions occupy the rightmost real-space sites (row $l=36$ here).

Generally, the total number of rows in the FS lattice is $1+N_{e}(N-N_{e}) = 1+\nu(1-\nu)N^{2}$  
(as the minimum number of sequential hops required to connect $|L\rangle$ to $|R\rangle$ is $N_{e}(N-N_{e})$). 
Since the FS lattice is invariant to reading the fermion strings in FS sites from left to right, or vice versa,
the number of FS sites in rows $l$ and $N_{e}(N-N_{e}) -l$ are the same.
The number of sites in row $l$ can be shown to be the number of integer partitions 
of $l$, subject to the restriction that no partition may have more than $N_{e}$ parts and no individual part can be larger than $N-N_{e}$; and asymptotically grows exponentially  in $\sqrt{l}$ as $l$ is increased towards the middle rows 
$l\sim\tfrac{1}{2}\nu(1-\nu)N^{2}$.

Although our focus here is on finite filling $\nu =N_{e}/N$ in the thermodynamic limit $N\rightarrow \infty$, for which 
the FS lattice clearly `balloons', we add that in the single-fermion limit $N_{e}=1$ it reduces simply to that for
the real-space lattice; with $N$ rows,   each containing just one state (in which the fermion occupies a particular 
real-space site).

Several general features of the FS lattice are apparent. First, the hopping $T_{IJ}$ is local  in state-space, in 
the sense that the FS sites $\{|J\rangle \}$ to which any given $|I\rangle$ connects under $H_{t}$ necessarily 
lie in a directly adjacent row, reflecting that one fermion has hopped to a NN site in the physical lattice.
Moreover, since the hypercubic real-space lattices considered are bipartite, it follows that the corresponding FS lattice 
is also bipartite.

Second, physical properties of the system are typically characterised by probability distributions -- over FS lattice 
sites, disorder realisations, or both. An obvious example is the local connectivity under  $T_{IJ}$, reflected in the distribution  of coordination numbers $Z_{I}$ over FS sites (which is independent of disorder). In e.g.\ a $1d$ chain the minimum  $Z_{I}$ for all fillings $\nu =N_{e}/N$ is clearly $Z_{I,\mathrm{min}} =1$ (or $2$ for periodic boundary conditions), occurring solely for the apical sites of rows $l=0$ and $N_{e}(N-N_{e})$ where the fermions are maximally bunched.
The maximum $Z_{I}$ by contrast is  macroscopically large, $Z_{I,\mathrm{max}}=2N_{e}$ for all fillings $\nu <1/2$, arising
for states in which each fermion is surrounded by at least two empty sites; with $Z_{I,\mathrm{max}}=2(N-N_{e})$ for $\nu >1/2$
 (and $Z_{I,\mathrm{max}} =N-1$ precisely at half-filling). 
 
The FS site energies (eq.\ \ref{eq:3}) are also naturally distributed, over sites and/or disorder realisations. 
These $\{\eI\}$ are the counterparts, for the equivalent Fock-space TBM  eq.\ \ref{eq:4}, of the site-energies 
$\{\epsilon_{i}\}$ in a one-particle TBM. One thus expects their distribution to influence whether many-body states are FS localized or extended; while recognising that, unlike the ($N$) real-space $\{\epsilon_{i}\}$ which are independent random variables, the $N_{\cal{H}}$ FS site-energies $\{\eI\}$ are correlated.  A further obvious example  is the eigenvalue spectrum of $H$ -- or density of states (DoS) -- and its distribution over disorder realisations.


\section{Distributions}
\label{section:distributions}

We seek then the distributions, over FS lattice sites and disorder realisations, of:
the FS site-energies $\{{\cal{E}}_{I}\}$, the eigenvalues of $H$,  the coordination numbers $Z_{I}$, and
$r_{I} =\sum_{\langle ij \rangle}\n{i}\n{j}$ of eq.\ \ref{eq:3} (which determines the interaction contribution to $\eI$). 

In the thermodynamic limit of interest we take it as given that all such distributions are Gaussian
(as is essentially obvious from the central limit theorem, although can be shown explicitly).
 We thus focus on first and second moments. As will be seen, all such may be obtained solely from a knowledge of 
that for $r_{I}$.

For any quantity $O_{I} =\langle I |\hat{O}|I\rangle$, its average $\overline{O}$ over both FS sites and disorder realisations  is
\begin{equation}
\label{eq:5}
\overline{O}~=~ \langle \mathrm{Tr}O_{I}^{\pd}\rangle_{\epsilon}^{\pd} .
\end{equation}
Here $\langle ....\rangle_{\epsilon} =\int \prod_{i=1}^{N} [d\epsilon_{i}P(\epsilon_{i})] ....$ denotes the disorder average, 
and $\mathrm{Tr}O_{I}$ an average over the $N_{{\cal{H}}} ={^{N}}C_{N_{e}}$ FS  sites/states
\begin{equation}
\label{eq:6}
\mathrm{Tr}O_{I}^{\pd}~=~N_{{\cal{H}}}^{-1} \sum_{I}~O_{I}^{\pd} ~=~N_{{\cal{H}}}^{-1}\sum_{I} \langle I|\hat{O}|I\rangle 
\end{equation}
(such that averages over disorder and FS sites commute, 
$\langle \mathrm{Tr}O_{I}\rangle_{\epsilon}\equiv \mathrm{Tr}\langle O_{I}\rangle_{\epsilon}$).

Consider a generic product $(\n{i}\n{j}\n{k} .....)$ corresponding to $m$ distinct real-space sites $i,j,k...$. 
This is non-zero only if all $m$ sites are occupied (occupation numbers of $1$), so $\sum_{I}(\n{i}\n{j}\n{k} .....)$ is simply the number of ways of distributing $(N_{e}-m)$ fermions over $(N-m)$ sites.
Hence $\mathrm{Tr}(\n{i}\n{j}\n{k} ...)=N_{{\cal{H}}}^{-1}\times{^{(N-m)}}C_{(N_{e}-m)} \equiv \nu_{m}$ is 
\begin{equation}
\label{eq:7}
\mathrm{Tr}(\n{i}\n{j}\n{k} .....)=\nu_{m}^{\pd}
=\prod_{n=0}^{m-1}\frac{N_{e}-n}{N-n} = \nu~\prod_{n=1}^{m-1}\frac{\nu-\frac{n}{N}}{1-\frac{n}{N}} 
\end{equation}
such that $\nu_{1} =\nu$, the filling fraction. Note that $\nu_{m} \equiv \nu^{m}+{\cal{O}}(1/N)$ in the thermodynamic limit 
(although eq.\ \ref{eq:7} holds for any $N, N_{e}$, and is required below).

Consider first the FS site-energies ${\cal{E}}_{I}=\langle I|H_{0}|I\rangle$ (eq.\ \ref{eq:3}).
$\mathrm{Tr}{\cal{E}}_{I}$ is the mean of ${\cal{E}}_{I}$ over FS sites for a given disorder realisation,
given (via eq.\ \ref{eq:7}) by
\begin{equation}
\label{eq:8}
\mathrm{Tr}{\cal{E}}_{I}^{\pd}~=~\nu \sum_{i}\epsilon_{i} ~+~V\overline{r}, 
\end{equation}
with $\overline{r} \equiv \mathrm{Tr}r_{I}$ (as $r_{I}$ is independent of disorder). Since the disorder-averaged 
$\langle \epsilon \rangle_{\epsilon}=0$, 
$\overline{\cal{E}}=\langle\mathrm{Tr}{\cal{E}}_{I}\rangle_{\epsilon} = V\overline{r}$.
From eqs.\ \ref{eq:3},\ref{eq:7} 
$\mathrm{Tr}r_{I} =\nu_{2}\sum_{\langle ij \rangle} =\nu_{2}Z_{\mathrm{d}}N/2$, so 
$\overline{\cal{E}}$ is given in the thermodynamic limit by
\begin{equation}
\label{eq:9}
\overline{\cal{E}}~=~V \overline{r}~=~V ~\tfrac{1}{2} Z_{\mathrm{d}} \nu^{2} N ~=~
V~\nu d N_{e} .
\end{equation}
Disorder-induced fluctuations in $\mathrm{Tr}{\cal{E}}_{I}$ are embodied
in $\langle [\mathrm{Tr}{\cal{E}}_{I}]^{2}\rangle_{\epsilon}$, and since
the lattice site-energies $\{\epsilon_{i}\}$ are independent random variables, 
$\langle \epsilon_{i}\epsilon_{j}\rangle_{\epsilon}^{\pd} =\delta_{ij}\langle\epsilon^{2}\rangle$ (with 
$\langle\epsilon^{2}\rangle \equiv \langle\epsilon^{2}\rangle_{\epsilon}^{\pd}$ for brevity); so
\begin{equation}
\label{eq:10}
\langle [\mathrm{Tr}{\cal{E}}_{I}^{\pd}]^{2}\rangle_{\epsilon}~=~\nu^{2}\langle\epsilon^{2}\rangle N ~+~V^{2}\overline{r}^{2} .
\end{equation}
$\overline{{\cal{E}}^{2}} =\langle\mathrm{Tr}({\cal{E}}_{I}^{2})\rangle_{\epsilon}$ likewise follows using 
eqs.\ \ref{eq:3},\ref{eq:7} as
\begin{equation}
\label{eq:11}
 \langle\mathrm{Tr}{\cal{E}}_{I}^{2}\rangle_{\epsilon}~=~ \overline{{\cal{E}}^{2}} ~=~\nu\langle\epsilon^{2}\rangle N
~+~V^{2}\overline{r^{2}}
\end{equation}
(where $\overline{r^{2}} =\mathrm{Tr}r_{I}^{2}$). For reasons explained in sec.\ \ref{section:sec4} we now define two distinct variances, specifically
$\mu_{\ecal}^{2} = \langle \mathrm{Tr}([\eI -\mathrm{Tr}\eI]^{2})\rangle_{\epsilon}$
and $\mu_{\ecal_{0}}^{2} = \langle \mathrm{Tr}([\eI -\langle\mathrm{Tr}\eI\rangle_{\epsilon}]^{2})\rangle_{\epsilon}$
$=\langle \mathrm{Tr}([\eI -\overline{\ecal}]^{2})\rangle_{\epsilon}$. These are thus given by
\begin{subequations}
\label{eq:12}
\begin{align}
\mu_{{\cal{E}}}^{2}=& \langle\mathrm{Tr}{\cal{E}}_{I}^{2}\rangle_{\epsilon}^{\pd} -
\langle [\mathrm{Tr}{\cal{E}}_{I}^{\pd}]^{2}\rangle_{\epsilon}
= \nu(1-\nu)\langle\epsilon^{2}\rangle N ~+~V^{2}\mu_{r}^{2}
\label{eq:12a}
\\
\mu_{{\cal{E}}_{0}}^{2}=&\langle\mathrm{Tr}{\cal{E}}_{I}^{2}\rangle_{\epsilon}^{\pd} -
\langle \mathrm{Tr}{\cal{E}}_{I}^{\pd}\rangle_{\epsilon}^{2}
=\nu\langle\epsilon^{2}\rangle N ~+~V^{2}\mu_{r}^{2}
\label{eq:12b}
\end{align}
\end{subequations}
with $\mu_{r}^{2} =\overline{r^{2}}-\overline{r}^{2}$ the variance of $r_{I}$.

We turn now to the distribution of eigenvalues. $\mathrm{Tr}H$ does not depend on the basis chosen, so
$\mathrm{Tr}H =N_{{\cal{H}}}^{-1}\sum_{I}\langle I|H|I\rangle$$\equiv N_{{\cal{H}}}^{-1}\sum_{n} E_{n}$, 
and gives the centre of gravity of the eigenvalues $\{ E_{n}\}$ for any given disorder realisation.
Since $\mathrm{Tr}H = \mathrm{Tr}{\cal{E}}_{I}$ (eq.\ \ref{eq:4}), the disorder-averaged mean eigenvalue
$\overline{E} = \langle\mathrm{Tr}H\rangle_{\epsilon}^{\pd}=\overline{{\cal{E}}}$, i.e.\
\begin{equation}
\label{eq:13}
\overline{E}~=~\overline{{\cal{E}}} ~=~V\overline{r}~=~V~\nu d N_{e}
\end{equation}
(with $\overline{E}\propto N_{e}$ reflecting extensivity). 
Since $\langle [\mathrm{Tr}H]^{2}\rangle_{\epsilon}=\langle[\mathrm{Tr}{\cal{E}}_{I}]^{2}\rangle_{\epsilon}$,
eq.\ \ref{eq:10} gives
\begin{equation}
\label{eq:14}
\langle \big[\mathrm{Tr}H\big]^{2}\rangle_{\epsilon} ~=~\nu^{2}\langle\epsilon^{2}\rangle N ~+~V^{2}\overline{r}^{2} .
\end{equation}
Now consider $\overline{E^{2}}=\langle\mathrm{Tr}H^{2}\rangle_{\epsilon}^{\pd}$.
From eq.\ \ref{eq:4}, $\langle I|H^{2}|I\rangle = {\cal{E}}_{I}^{2} +\sum_{J}T_{IJ}T_{JI}$
$={\cal{E}}_{I}^{2} +t^{2}Z_{I}$ with $Z_{I}$ the coordination number of FS site $I$. Hence 
\begin{equation}
\label{eq:15}
\langle\mathrm{Tr}H^{2}\rangle_{\epsilon}^{\pd}~=~\overline{{\cal{E}}^{2}}~+~t^{2}\overline{Z}
\end{equation}
with  $\overline{Z}=\mathrm{Tr}Z_{I}$ the average coordination number of the FS lattice.
 
In parallel to the FS site energies above, we define again two variances, 
$\mu_{E}^{2} = \langle \mathrm{Tr}([H-\mathrm{Tr}H]^{2})\rangle_{\epsilon}$ and
$\mu_{E_{0}}^{2} =\langle \mathrm{Tr}([H-\langle\mathrm{Tr}H\rangle_{\epsilon}]^{2})\rangle_{\epsilon}$
$=\langle \mathrm{Tr}([H-\overline{E}]^{2})\rangle_{\epsilon}$; which  follow as
\begin{subequations}
\label{eq:16}
\begin{align}
\mu_{E}^{2}~=~ &\langle \mathrm{Tr}\big([H-\mathrm{Tr}H]^{2}\big)\rangle_{\epsilon}
\nonumber
\\
=~&\nu(1-\nu)\langle\epsilon^{2}\rangle N~+~V^{2}\mu_{r}^{2}~+~t^{2}\overline{Z}
\label{eq:16a}
\\
\mu_{E_{0}}^{2}=~&\langle \mathrm{Tr}\big([H-\langle\mathrm{Tr}H\rangle_{\epsilon}]^{2}\big)\rangle_{\epsilon}
\nonumber
\\
=~&\nu \langle\epsilon^{2}\rangle N ~+~V^{2}\mu_{r}^{2}~+~t^{2}\overline{Z}
\label{eq:16b}.
\end{align}
\end{subequations}
Physically, $\mu_{E}^{2}$  gives the disorder-averaged variance of the eigenvalues, relative to their centre of gravity 
$\mathrm{Tr}H=N_{{\cal{H}}}^{-1}\sum_{n}E_{n}$ for \emph{each} disorder realisation;  while $\mu_{E_{0}}^{2}$ 
gives the variance  relative to the full mean $\langle \mathrm{Tr}H\rangle_{\epsilon}$ over both FS sites and disorder. The reasons for introducing these two distinct variances are discussed in sec.\ \ref{section:sec4}. Here we simply note that it is $\mu_{E}^{2}$ (and likewise $\mu_{{\cal{E}}}^{2}$) which is of primary relevance.

The variances in eqs.\ \ref{eq:12},\ref{eq:16} thus follow once $\mu_{r}^{2}$ and the mean FS coordination number 
$\overline{Z}$ are known.  Obviously neither of the latter  depends on disorder. Further, as now shown, the coordination number $Z_{I}$ is simply related to $r_{I}$, so only the latter need be considered.

Each configuration of fermions on the real-space lattice specifies uniquely one FS basis state $|I\rangle$. With $1$ denoting an occupied site and $0$ an empty site, the following types of NN pairs arise: $11$, $10$ and $00$.
The \emph{total} number of such pairs in any $|I\rangle$ are denoted $\Ntot{mn}$, and are clearly 
\begin{subequations}
\label{eq:17}
\begin{align}
\Ntot{11} ~=&~\sum_{\langle ij\rangle} \n{i}\n{j}~~~ =~ r_{I}^{\pd}
\label{eq:17a}
\\
\Ntot{10}~=&~\sum_{\langle ij\rangle} \big[\n{i}(1-\n{j}) + (1-\n{i})\n{j}\big] =Z_{I}^{\pd}
\label{eq:17b}
\\
\Ntot{00}~=&~\sum_{\langle ij\rangle}(1-\n{i})(1-\n{j}) .
\label{eq:17c}
\end{align}
\end{subequations}
Using $\sum_{\langle ij\rangle} = Z_{\mathrm{d}}N/2$,  eqs.\ \ref{eq:17} naturally sum to the total number of NN pairs, 
$\Ntot{11}+\Ntot{10}+\Ntot{00} = \tfrac{1}{2}NZ_{\mathrm{d}} = dN$. Obviously $\Ntot{11} = r_{I}$ (see eq.\ \ref{eq:3}). Equally obviously $\Ntot{10}=Z_{I}$, the coordination number of FS site $I$, because each $10$-pair enables a fermion to hop under $t$; and using  
$\sum_{\langle ij\rangle}\n{i} = \tfrac{1}{2}Z_{\mathrm{d}}\sum_{i}\n{i} = \tfrac{1}{2}Z_{\mathrm{d}}N_{e}$,
eq.\ \ref{eq:17b} gives
\begin{equation}
\label{eq:18}
Z_{I}^{\pd}~=~ Z_{\mathrm{d}}N_{e} - 2r_{I}^{\pd} ~=~2(dN_{e} -r_{I}^{\pd}).
\end{equation}
The distribution of FS coordination numbers thus follows directly from a knowledge of that for $r_{I}$. Eq.\ \ref{eq:18} is also physically clear: for $r_{I} =\Ntot{11} =0$, all NN sites to each occupied site in $|I\rangle$ are empty, so each of the 
$N_{e}$ fermions can hop under $t$ to its $2d$ NNs, whence $Z_{I} =2dN_{e}$; while for $r_{I} =\Ntot{11} \neq 0$,
 each additional NN $11$-pair clearly `blocks' one hop for each of the pair of fermions, and thus reduces $Z_{I}$ by $2$.


\subsection{Variance of $r_{I}^{~}$}
\label{subsection:sec3a}

From eqs.\ \ref{eq:12},\ref{eq:16} the final step is to determine the second moment 
$\overline{r^{2}}=\mathrm{Tr}\sum_{\langle ij \rangle}\sum_{\langle kl \rangle} \n{i}\n{j}\n{k}\n{l}$.
Since the sites in $\langle ij\rangle$ and $\langle kl\rangle$ are not all distinct we partition this sum into terms involving $2$ NN sites, $3$ adjacent NN sites, and two pairs of distinct NN sites, specifically
\begin{equation}
\label{eq:19}
\overline{r^{2}}~=~\sum_{\langle ij \rangle} \nu_{2}^{\pd} ~+~2\sum_{\langle ij\rangle}\sum_{k}\nu_{3}^{\pd}
~+~ {\sum_{\langle ij\rangle,\langle kl\rangle}}^{\prime}\nu_{4}^{\pd} .
\end{equation}
The middle sum refers to $3$ distinct sites,
where the site $k \neq i$ or $j$, but is a NN to one of them, such that for any given $\langle ij\rangle$ there are $Z_{\mathrm{d}}-1$ sites in the $k$ sum.
Hence, $2\sum_{\langle ij\rangle}\sum_{k}\nu_{3}^{\pd} =2(Z_{\mathrm{d}}-1)\nu_{3}\sum_{\langle ij\rangle}$
$= NZ_{\mathrm{d}}(Z_{\mathrm{d}}-1)\nu_{3}$. The final sum in eq.\ \ref{eq:19} refers to two pairs of distinct NN sites.
Recognising that the original sum for $\overline{r^{2}}$ contained a total of $[\tfrac{1}{2}Z_{\mathrm{d}}N]^{2}$ terms, while the first pair of terms on the right side of eq.\ \ref{eq:19} contain respectively $\tfrac{1}{2}Z_{\mathrm{d}}N$ and
$NZ_{\mathrm{d}}(Z_{\mathrm{d}}-1)$ terms, gives 
$\sum^{\prime}_{\langle ij\rangle,\langle kl\rangle}\nu_{4} =\tfrac{1}{2}Z_{\mathrm{d}}N \big[
\tfrac{1}{2}Z_{\mathrm{d}}N -(2Z_{\mathrm{d}}-1)\big]\nu_{4}$. 
Using the precise forms for $\nu_{m}$ (eq.\ \ref{eq:7}), together with 
$\overline{r} =\tfrac{1}{2}Z_{\mathrm{d}}N \nu_{2}$, then gives
$\mu_{r}^{2} =\overline{r^{2}} -\overline{r}^{2}$ as
\begin{equation}
\label{eq:20}
\mu_{r}^{2}=
\frac{\frac{1}{2}Z_{\mathrm{d}}\nu(1-\nu)N^{2}\big(N-[Z_{\mathrm{d}}+1]\big)(N_{e}-1)(N-N_{e}-1)}{(N-1)^{2}(N-2)(N-3)}.
\end{equation}
This is exact under periodic boundary conditions for finite $N$. In the thermodynamic limit of interest it gives the desired result for the variance of $r_{I}$,
\begin{equation}
\label{eq:21}
\mu_{r}^{2}~=~\nu^{2}(1-\nu)^{2}\tfrac{1}{2}Z_{\mathrm{d}}N
~=~ \nu(1-\nu)^{2} dN_{e} 
\end{equation}
(with leading corrections ${\cal{O}}(1)$).\\

Let us simply recap  the essential results arising in the thermodynamic limit. Eq.\ \ref{eq:13} gives the mean energies, 
$\overline{E}=\overline{{\cal{E}}} = V\overline{r} = V\nu d N_{e}$. For the variances, eqs.\ \ref{eq:12},\ref{eq:16} give
\begin{subequations}
\label{eq:22}
\begin{align}
\mu_{E}^{2}~=&~\mu_{{\cal{E}}}^{2}~+~ t^{2}\overline{Z}
\label{eq:22a}
\\
\mu_{{\cal{E}}}^{2}~=&~\mu_{W}^{2}~+~V^{2} \mu_{r}^{2} ~~~~:~\mu_{W}^{2} = 
(1-\nu)\langle \epsilon^{2}\rangle N_{e} 
\label{eq:22b}
\end{align}
\end{subequations}
and likewise 
\begin{subequations}
\label{eq:23}
\begin{align}
\mu_{E_{0}}^{2}~=&~\mu_{{\cal{E}}_{0}}^{2}~+~ t^{2}\overline{Z}
\label{eq:23a}
\\
\mu_{{\cal{E}}_{0}}^{2}~=&~\mu_{W_{0}}^{2}~+~V^{2} \mu_{r}^{2} ~~~~:~\mu_{W_{0}}^{2} = \langle \epsilon^{2}\rangle N_{e}~,
\label{eq:23b}
\end{align}
\end{subequations}
all being a sum of independent contributions from disorder, interactions and (for the eigenvalues) hopping.
The mean FS coordination number follows directly from eq.\ \ref{eq:18} as
$\overline{Z} = Z_{\mathrm{d}}N_{e}-2\overline{r}$, so from eq.\ \ref{eq:9} for $\overline{r}$
\begin{equation}
\label{eq:24}
\overline{Z}~=~ Z_{\mathrm{d}}(1-\nu) N_{e}
~=~2\nu(1-\nu) d N ;
\end{equation}
while the variance $\mu_{Z}^{2}=4\mu_{r}^{2}$ (from eq.\ \ref{eq:18}), with $\mu_{r}^{2}$ given by eq.\ \ref{eq:21}.
All these quantities follow directly on specifying  the filling fraction $\nu =N_{e}/N$, the space dimension $d$ of 
the real-space lattice, and the number of fermions $N_{e}$. 
Note also that $\mu_{E}^{2}$ and $\mu_{{\cal{E}}}^{2}$ are invariant under a particle-hole 
transformation $\nu\leftrightarrow 1-\nu$ (since $\mu_{W}^{2} =\nu(1-\nu)\langle\epsilon^{2}\rangle N$);
$\mu_{E_{0}}^{2}$ and $\mu_{{\cal{E}}_{0}}^{2}$ by contrast are not.

\begin{figure}
\includegraphics{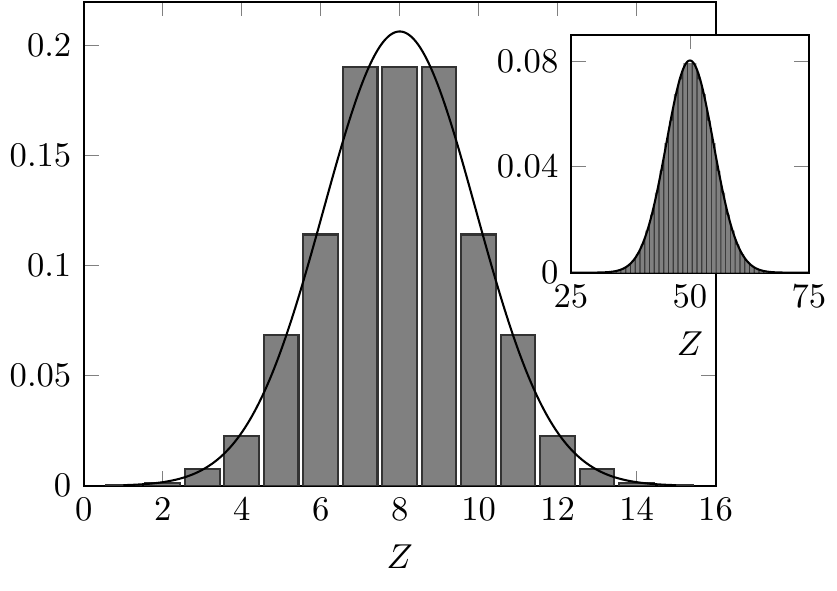}
\caption{\label{fig:fig2} 
Distribution $p(N,N_{e}; Z)$ of coordination numbers $Z$ for the FS lattice arising for the
$1d$ chain, shown for $N=16, N_{e}=8$; and compared to the normal distribution (solid line) of mean $\overline{Z}$ 
(eq.\ \ref{eq:24}) and variance $\mu_{Z}^{2}=4\mu_{r}^{2}$ (eq.\ \ref{eq:21}) appropriate to the thermodynamic limit. 
\emph{Inset}: same comparison for $N=100, N_{e}=50$.
}
\end{figure}

~\\
The distribution of FS coordination numbers (or equivalently $r_{I}$, eq.\ \ref{eq:18}) itself depends solely on the lattice, and not on $H$ or its parameters. For the $1d$ chain, this distribution can be determined for arbitrary $N, N_{e}$ from basic combinatorics. With $p(N,N_{e};Z)$ denoting the fraction of FS lattice sites with coordination number $Z$, it is given by 
(explicitly here for an open chain): 
\begin{equation}
\nonumber
\begin{split}
N_{{\cal{H}}}^{\pd}&p(N,N_{e}; Z)
\\
~=~&2~~\tfrac{(N-Z)}{Z}~{^{(N_{e}-1)}}C_{\frac{Z}{2}-1}~{^{(N-N_{e}-1)}}C_{\frac{Z}{2}-1}
~~~~:~Z~\mathrm{even}
\\
~=~&~2~~{^{(N_{e}-1)}}C_{\frac{Z}{2}-\frac{1}{2}}~{^{(N-N_{e}-1)}}C_{\frac{Z}{2}-\frac{1}{2}}
~~~~~~~~~~~:~Z~\mathrm{odd}
\end{split}
\end{equation}
As an indication of the size-dependence of this distribution,
Fig.\ \ref{fig:fig2} shows $p(N,N_{e};Z)$ \emph{vs} $Z$ for a half-filled $N=16$-site system (as typically 
employed in exact diagonalisation studies of MBL, and with $N_{{\cal{H}}}=12870$ FS sites).
The Gaussian distribution appropriate to the thermodynamic limit is shown for comparison.
Even for $N=16$ this is quite well approached, despite the modest range of available $Z$'s; and 
by $N=100$ (fig.\ \ref{fig:fig2} inset) the discrete and normal distributions are barely distinguishable. 
Analogous comparison for the eigenvalue distribution is given below.

\section{Energy Variances}
\label{section:sec4}

We have introduced two distinct variances for the eigenvalues (eqs.\ \ref{eq:16}), and the FS site-energies 
(eq.\ \ref{eq:12}). Here we explain why, and why it is that $\mu_{E}^{2}$ and  $\mu_{{\cal{E}}}^{2}$ are the 
physically relevant variances; focusing  in the following on the eigenvalues (the same considerations apply 
to the FS site-energies). The reason is  a little subtle, does not usually arise in considering one-body 
localization (1BL) -- although it may do, as explained -- and has implications for the identification of 
mobility edges. 

Consider first the 1BL case of a single fermion,~\cite{FN4} with filling $\nu =1/N$. Here, trivially, 
$\overline{Z} \equiv Z_{\mathrm{d}}$ is the coordination number of the physical lattice   and 
$\overline{r}=0 =\mu_{r}^{2}$ (since $N_{e}=1$). In this case (eq.\ \ref{eq:16}), $\mu_{E}^{2} \equiv \mu_{E_{0}}^{2}$ 
($=\langle\epsilon^{2}\rangle +t^{2}Z_{\mathrm{d}}$) coincide in the thermodynamic limit, so which of the two one considers 
is immaterial. This reflects the fact (eq.\ \ref{eq:14}) that for 1BL
$\langle [\mathrm{Tr}H]^{2}\rangle_{\epsilon} =\langle\epsilon^{2}\rangle/N$ vanishes as 
$N\rightarrow \infty$, i.e.\ that $\mathrm{Tr}H =N_{{\cal{H}}}^{-1}\sum_{n}E_{n}$ -- the 
centre of gravity of the eigenvalue distribution for any given disorder realisation -- 
does not fluctuate from realisation to realisation. For a macroscopic system, each realisation will then 
yield the same eigenvalue spectrum (self-averaging). By itself, that spectrum does not of course contain information about whether states are localized (L) or extended (E). But states of any given energy are either L or E with probability unity over an ensemble of disorder realisations; so the fact that the same spectrum is obtained for all disorder realisations ensures 
a pristine distinction between L and E states as a function of energy, and hence an unambiguous identification of mobility edges separating them.

The situation above is not however ubiquitous, even for 1BL. To illustrate this, consider the Hamiltonian
\begin{subequations}
\label{eq:25}
\begin{align}
H^{\prime}=~&\sum_{i} \epsilon_{i}^{\pd} (\hat{n}_{i}^{\pd}-\tfrac{1}{2}) +
H_{t}+\sum_{\langle ij \rangle} V(\hat{n}_{i}^{\pd}-\tfrac{1}{2})(\hat{n}_{j}^{\pd}-\tfrac{1}{2})
\label{eq:25a}
\\
=~&c +\sum_{i} \epsilon_{i}^{\pd} (\hat{n}_{i}^{\pd}-\tfrac{1}{2})+
H_{t}
+\sum_{\langle ij \rangle} V~\hat{n}_{i}^{\pd}\hat{n}_{j}^{\pd}
\label{eq:25b}
\end{align}
\end{subequations}
(with $c = \tfrac{1}{2}VZ_{\mathrm{d}}(\tfrac{1}{4}-\nu)N$  a constant which is irrelevant in the following).
As for the $H$ of eq.\ \ref{eq:1}, this Hamiltonian is widely studied in MBL, since in $d=1$ it maps directly to a
random $XXZ$ model under a Jordan-Wigner transformation. $H^{\prime}$ is regarded as being equivalent to $H$; and 
indeed, aside from the trivial disorder-independent constant ($c$), the two Hamiltonians  differ 
only by a constant:  $H^{\prime} \equiv H-\tfrac{1}{2}C$ with $C= \sum_{i}\epsilon_{i}$.  
However $C$ depends on the disorder realisation, $C \equiv C(\{\epsilon_{i}\})$;
with a vanishing disorder-averaged mean $\overline{C} =0$, but a non-zero variance
$\overline{C^{2}} = \langle \epsilon^{2}\rangle N$ proportional to system size $N$.
For $H^{\prime}$ one has $\mathrm{Tr}H^{\prime} = (\nu-\tfrac{1}{2})\sum_{i}\epsilon_{i} +V\overline{r}$
$=(\nu-\tfrac{1}{2})C +V\overline{r}$, with the variances $\mu_{E}^{2}$ and $\mu_{E_{0}}^{2}$ given by
\begin{subequations}
\label{eq:26}
\begin{align}
\mu_{E}^{2}~=&~ 
\nu(1-\nu)\langle\epsilon^{2}\rangle N~+~V^{2}\mu_{r}^{2}~+~t^{2}\overline{Z}
\label{eq:26a}
\\
\mu_{E_{0}}^{2}~=&
\tfrac{1}{4} \langle\epsilon^{2}\rangle N ~+~V^{2}\mu_{r}^{2}~+~t^{2}\overline{Z} .
\label{eq:26b}
\end{align}
\end{subequations}
Note that  $\mu_{E}^{2} = \langle \mathrm{Tr}([H^{\prime}-\mathrm{Tr}H^{\prime}]^{2})\rangle_{\epsilon}$ for 
$H^{\prime}$ is identical to that arising for $H$ (eq.\ \ref{eq:16a}), so is unaffected by $C \equiv C(\{\epsilon_{i}\})$; 
by contrast, 
$\mu_{E_{0}}^{2} =\langle \mathrm{Tr}([H^{\prime}-\langle\mathrm{Tr}H^{\prime}\rangle_{\epsilon}]^{2})\rangle_{\epsilon}$
differs from that for $H$ (eq.\ \ref{eq:16b}).

\begin{figure}
\includegraphics{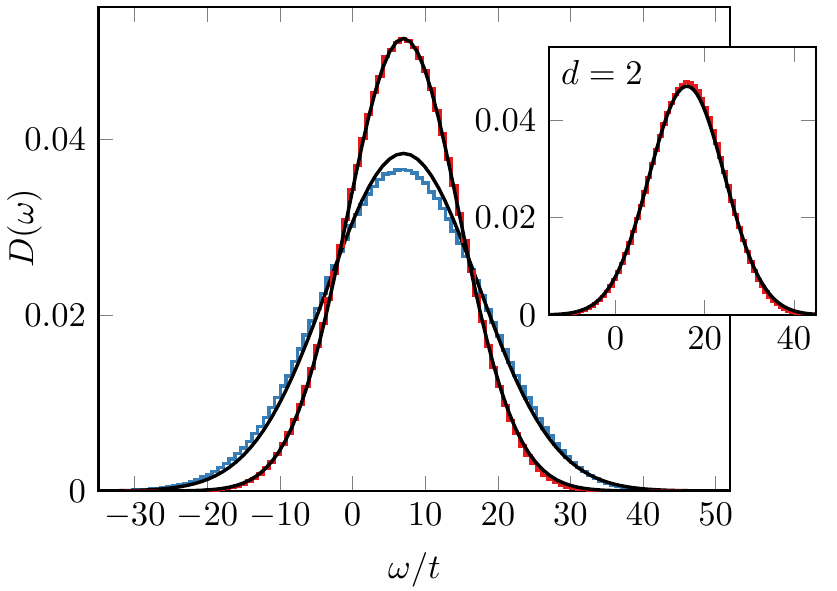}
\caption{\label{fig:fig3} 
Many-body eigenvalue spectrum $D(\w)$ \emph{vs} $\w/t$ (red), with site-energy distribution 
$P(\epsilon)=\tfrac{1}{W}\theta(\tfrac{1}{2}W -|\epsilon|)$, shown for $W/t=12, V/t =2$ at half-filling 
$\nu =\tfrac{1}{2}$; obtained numerically for an $N=16$-site chain ($200$ disorder realisations), and compared 
to the Gaussian thermodynamic limit result eq.\ \ref{eq:27} with variance $\mu_{E}^{2}$ of eq.\ \ref{eq:16a} (black line).
The corresponding `fully-averaged' spectrum of variance $\mu_{E_{0}}^{2}$ discussed in text is also shown (blue).
\emph{Inset}: $D(\w)$ \emph{vs} $\w/t$ (red) for an $N=4\times 4$ square lattice (with same parameters as for the
$d=1$ case), again compared to the Gaussian eq.\ \ref{eq:27} with variance $\mu_{E}^{2}$ (black line).
}
\end{figure}

We return to this below, but first consider again the 1BL limit, where $\nu =1/N$.
In this case, unlike that for $H$ considered above, $\mu_{E}^{2}$ and $\mu_{E_{0}}^{2}$ 
(eqs.\ \ref{eq:26}) no longer coincide:
$\mu_{E}^{2}=\langle\epsilon^{2}\rangle +t^{2}Z_{\mathrm{d}}$ (as arises also for $H$), while 
$\mu_{E_{0}}^{2}=\tfrac{1}{4} \langle\epsilon^{2}\rangle N +t^{2}Z_{\mathrm{d}}$ grows with system size $N$.
This reflects the fact that the centre of gravity of the eigenvalue distribution 
($\mathrm{Tr}H^{\prime}$) fluctuates from realisation to realisation,
$\langle [\mathrm{Tr}H^{\prime}]^{2}\rangle_{\epsilon} = \tfrac{1}{4}\langle \epsilon^{2}\rangle N 
=\tfrac{1}{4}\overline{C^{2}}$ growing with $N$ (in contrast to 
$\langle [\mathrm{Tr}H]^{2}\rangle_{\epsilon} =\langle\epsilon^{2}\rangle/N$).
The implications of this are clear -- for a macroscopic system each disorder realisation no longer yields the same eigenvalue spectrum as a function of energy. Rather, any two disorder realisations will yield identical copies of the spectrum, but energetically displaced/offset from each other by the difference in the $C \equiv C(\{\epsilon_{i}\})$'s for
the two realisations. This merely reflects the fact that for any given disorder realisation, 
$H^{\prime} \equiv H-\tfrac{1}{2}C$ and $H$ have exactly the same eigen\emph{states}, which hence have the same L or E character; but their eigen\emph{values}, while in $1$:$1$ correspondence, are each shifted by the disorder-dependent 
$C(\{\epsilon_{i}\})$. The cure is obvious: to restore the sharp distinction between L and E states as a function of energy,
as required for unambiguous identification of mobility edges, one needs only to eliminate these realisation-dependent offsets; thereby referring all energies to a common origin independent of disorder 
(specifically $\langle\mathrm{Tr}H^{\prime}\rangle_{\epsilon} =0=\langle\mathrm{Tr}H\rangle_{\epsilon}$), 
with eigenvalue fluctuations treated relative to their centre of gravity for \emph{each} disorder realisation.
It is of course precisely this which is captured by 
$\mu_{E}^{2}=\langle \mathrm{Tr}([H^{\prime}-\mathrm{Tr}H^{\prime}]^{2})\rangle_{\epsilon}$
($=\langle \mathrm{Tr}([H-\mathrm{Tr}H]^{2})\rangle_{\epsilon}$), hence our focus on it
rather than on $\mu_{E_{0}}^{2}$.

The 1BL situation just described for the case of $H^{\prime}$ is in fact the norm when considering MBL (where the filling 
$\nu =N_{e}/N$ is strictly non-vanishing in the thermodynamic limit). Here, whether $H$ or $H^{\prime}$ is considered, the centre of gravity of the eigenvalue distribution fluctuates with disorder realisation, with
 $\langle [\mathrm{Tr}H]^{2}\rangle_{\epsilon}=\nu^{2}\langle\epsilon^{2}\rangle N +V^{2}\overline{r}^{2}$ (eq.\ \ref{eq:14})
and $\langle [\mathrm{Tr}H^{\prime}]^{2}\rangle_{\epsilon}$ 
$=[\nu-\tfrac{1}{2}]^{2}\langle\epsilon^{2}\rangle N + V^{2}\overline{r}^{2}$ each inevitably $\propto N$.
The resolution of the disorder-induced offsets is precisely the same as for the 1BL example above:
all energies are referred to a common origin 
($\overline{E} =V\overline{r} =\langle\mathrm{Tr}H^{\prime}\rangle_{\epsilon} =\langle\mathrm{Tr}H\rangle_{\epsilon}$), 
with eigenvalue fluctuations treated relative to their centre of gravity ($\mathrm{Tr}H$ or $\mathrm{Tr}H^{\prime}$)
for each disorder realisation, as embodied in $\mu_{E}^{2}$.

With this, the (normalised) eigenvalue spectrum  $D(\w) = N_{{\cal{H}}}^{-1}\sum_{n}\delta(\w -E_{n})$
($\equiv N_{{\cal{H}}}^{-1}\langle\sum_{n}\delta(\w -E_{n})\rangle_{\epsilon}$) is given by the Gaussian
\begin{equation}
\label{eq:27}
D(\w) ~=~\frac{1}{\sqrt{2\pi}\mu_{E}}\exp \Big(- \frac{[\w -\overline{E}]^{2}}{2\mu_{E}^{2}}\Big).
\end{equation}
Note again that $\mu_{E}^{2}$ is the same for both Hamiltonians $H$ and $H^{\prime}$ (eqs.\ \ref{eq:16a},\ref{eq:26a}), 
whence so too are their spectra  $D(\w)$. Indeed this is readily seen to be true for any Hamiltonian with a site-energy term
$\sum_{i} \epsilon_{i} (\hat{n}_{i}-\zeta)$ with $\zeta$ an arbitrary disorder-independent constant, encompassing 
$H$ and $H^{\prime}$ as particular cases. It is eq.\ \ref{eq:27} we refer to in the following as the eigenvalue spectrum/DoS.

The previous considerations are salutary. If one does not account for disorder-induced energy offsets as above,
and instead averages the eigenvalue distribution willy nilly over all disorder realisations, then if a sharp distinction between L or E states as a function of energy occurs, it will be lost (as above). The resultant averaged  distribution in this case is again Gaussian with the same mean $\overline{E}=V\overline{r}$, but now with a variance
$\mu_{E_{0}}^{2} = \langle \mathrm{Tr}([H-\langle\mathrm{Tr}H\rangle_{\epsilon}]^{2})\rangle_{\epsilon}$
(with $\mu_{E_{0}}^{2}$ differing for $H$ and $H^{\prime}$, eqs.\ \ref{eq:16b},\ref{eq:26b} respectively, and 
$\mu_{E_{0}}^{2} \neq \mu_{E}^{2}$ for either $H$ or $H^{\prime}$).

These differences are clearly evident in finite-size calculations. We illustrate  them 
in fig.\ \ref{fig:fig3}, considering the Hamiltonian $H$ (eq.\ \ref{eq:1}) for the $d=1$ chain, with a standard box distribution $P(\epsilon)=\tfrac{1}{W}\theta(\tfrac{1}{2}W -|\epsilon|)$ for the site-energy distribution (and choosing 
$W/t =12$, $V/t=2$). Results are shown for $N=16$ sites at half-filling $\nu =\tfrac{1}{2}$, generated from $200$ disorder realisations. The resultant eigenvalue spectrum is shown [red],~\cite{FN5} and compared (solid line) to the Gaussian 
$D(\w)$ eq.\ \ref{eq:27} with variance $\mu_{E}^{2}$ (eq.\ \ref{eq:16a}). The latter is seen to be 
excellently captured, even for $N=16$. The fully-averaged spectrum is also shown [blue], and likewise compared to a Gaussian of variance  $\mu_{E_{0}}^{2}$ (eq.\ \ref{eq:16b}), which similarly captures it  well. The two spectra are visibly distinct 
(with $\mu_{E_{0}}^{2}>\mu_{E}^{2}$), as expected from the considerations above.

The inset to fig.\ \ref{fig:fig3} also shows the numerical DoS for a $d=2$-dimensional square lattice (dimensionality $d$ entering both the mean eigenvalue $\overline{E}$,  eq.\ \ref{eq:13}, and the variance eq.\ \ref{eq:16a} via the interaction and hopping terms, $V^{2}\mu_{r}^{2}$ and $t^{2}\overline{Z}$). Parameters considered are otherwise the same as those for 
$d=1$ in the main figure, and comparison is made to the Gaussian $D(\w)$ eq.\ \ref{eq:27} with variance eq.\ \ref{eq:16a}.  
Despite the relatively meagre $N=4\times 4$ real-space lattice, this again captures the numerics very well.

Finally, note that the  DoS eq.\ \ref{eq:27} is obviously $N$-dependent, in two ways: via $\overline{E} \propto N$ (which is trivially dealt with by referring energies relative to the band centre  $\w =\overline{E}$), and because its standard deviation $\mu_{E} \propto \sqrt{N}$. As such, it is natural to rescale energies  as $\wtil = (\w -\overline{E})/\mu_{E}$, such that the DoS $\tilde{D}(\wtil)$ (normalised to unity over $\wtil$)  is a standard normal distribution,
\begin{equation}
\label{eq:28}
\tilde{D}(\wtil)~=~\frac{1}{\sqrt{2\pi}}\exp\big(-\tfrac{1}{2}\wtil^{2}\big)
~~~~:~ \wtil =\frac{(\w-\overline{E})}{\mu_{E}} .
\end{equation}


\section{Conclusion}
\label{section:sec5}

We  have considered a canonical model employed in studies of MBL: a disordered system of interacting  spinless fermions, 
here  on a $d$-dimensional  lattice with $N \equiv L^{d}$ sites and $N_{e}$ fermions, for  non-vanishing filling 
$\nu =N_{e}/N$.  The model can be cast as an equivalent tight-binding model (eq.\ \ref{eq:4}) on a locally connected
Fock-space lattice of dimension $N_{{\cal{H}}}\propto e^{cN}$, the sites of which correspond to the  many-particle states 
of the system in the absence of hopping. As such, precisely the same questions may be asked as for a conventional one-body 
TBM, including about the density of many-body states, and whether those states are FS localized or extended.
We have barely touched on the latter question here; but, as a precursor to it, have considered the distributions 
-- over FS lattice sites and/or disorder realisations as appropriate -- of the system's eigenvalues, the FS site energies, 
and the local FS coordination numbers. All such are normally distributed, with variances in the thermodynamic limit that 
are readily determined, and found to be well captured by exact diagonalisation on the small system sizes of up to $N=16$ 
sites typically used in numerical work.

Some aspects of the results above warrant final  brief comment. First, from the discussion in sec.\ \ref{section:sec4} 
of the eigenvalue spectrum $D(\w)$, eq.\ \ref{eq:27}  (or $\tilde{D}(\wtil)$, eq.\ \ref{eq:28}), all but an exponentially 
small fraction of states lie in a `$\sqrt{N}$ scaling window' about the band centre $\w =\overline{E}$, i.e.\ on energy 
scales set by $\mu_{E} \propto \sqrt{N}$. It is worth considering what implications this might have for many-body mobility 
edges (ME). Above a certain non-zero critical disorder $W =W_{c}$, states at the band centre -- and by presumption all 
states -- are MBL. What then happens as $W$ is reduced below $W_{c}$?  Without prejudice there would  seem to be two 
distinct possibilities: either essentially all states become extended as $W$ is decreased just below $W_{c}$; or not.
Were the former to occur, we have nothing to say about it. But if the latter arises,  one expects many-body MEs at 
$[\w -\overline{E}] =\w_{\mathrm{mob}\pm}$ to open up  \emph{continuously} about the band centre, separating many-body 
localized  from extended states; as indeed detailed numerical work 
finds.~\cite{LuitzAletPRB2015,BeraFHMBardarsonPRL2015,Devakul15,Serbyn+P+AbaninPRX2015}
In this case we simply  point out that -- by virtue of the $\mu_{E} \propto \sqrt{N}$ scaling of the eigenvalue spectrum -- 
mobility edge trajectories (as MEs move further into the band with decreasing $W$) must likewise scale with $\sqrt{N}$, 
and not with $N$. While this does not preclude a subsequent crossover  to MEs scaling with $N$ itself, 
by definition the latter can occur only when $\w_{\mathrm{mob}}$ lies ${\cal{O}}(N)$ away from the band centre;
and as such lies deep in the tails of the eigenvalue spectrum, where the fraction of states is exponentially small.

These considerations have implications for numerically determined mobility edges. The energy axis 
in such studies is commonly expressed as an energy 
density~\cite{LuitzAletPRB2015,BeraFHMBardarsonPRL2015,Devakul15,Serbyn+P+AbaninPRX2015} 
$\overline{\w} =(\w -E_{\mathrm{min}})/(E_{\mathrm{max}}-E_{\mathrm{min}})$; with $E_{\mathrm{min/max}}$ 
the smallest/largest eigenvalue, and $\overline{\w}=1/2$ corresponding to the band centre. $E_{\mathrm{min}}$ and 
$E_{\mathrm{max}}$ are however each ${\cal{O}}(N)$ removed in energy from the band centre, with 
$E_{\mathrm{max}}-E_{\mathrm{min}} \propto N$. Hence if a  mobility edge at $\w$ lies within, say, a few 
$\mu_{E} \propto \sqrt{N}$ of the band centre, such that a fraction  ${\cal{O}}(1)$ of states are delocalised, then 
$\overline{\w} \rightarrow \tfrac{1}{2} +{\cal{O}}(1/\sqrt{N})$ nevertheless `sticks' at $1/2$ in the 
thermodynamic limit; and departs from $1/2$ only upon delocalisation 
of exponentially rare states lying deep in the spectral tails, outside the $\sqrt{N}$ scaling window.
If by contrast the energy axis is scaled in terms of $\mu_{E} \propto \sqrt{N}$, then the continuous evolution 
of mobility edges as they move through the eigenvalue spectrum will be captured.
Of course these considerations refer to the thermodynamic limit, and it is not \emph{a priori} clear whether the 
modest system sizes amenable to numerics would be sufficient in practice to distinguish between $\sqrt{N}$ and $N$ 
behaviour; although re-evaluation of previously obtained numerical data along the lines suggested here 
should shed light on the matter.

A further aspect of $\overline{\w} =(\w -E_{\mathrm{min}})/(E_{\mathrm{max}}-E_{\mathrm{min}})$
-- viz.\ that the band centre is identified as $\overline{\w} =1/2$, i.e.\ by 
$\w =\tfrac{1}{2}(E_{\mathrm{min}}+E_{\mathrm{max}})$ -- relates to the discussion of sec.\ \ref{section:sec4}.
Since $E_{\mathrm{min/max}}$ are by definition the extremal eigenvalues, they lie deep in the exponential tails 
of the eigenvalue spectrum, and will each fluctuate considerably from disorder realisation to realisation. 
Identifying the band centre by $\w =\tfrac{1}{2}(E_{\mathrm{min}}+E_{\mathrm{max}})$ will then blur
the pristine distinction between localised and delocalised states as a function of energy, required for
optimal identification of mobility edges. To circumvent this, it would be preferable 
(sec.\ \ref{section:sec4}) to identify the band centre for each disorder realisation from the centre of gravity 
$\mathrm{Tr}H$ of the eigenvalue spectrum, with energies referred relative to that natural `origin'.

Finally, we point out the obvious fact that  a typical local coordination number for the FS lattice at  non-zero 
filling  is its average, $\overline{Z} \propto N$ (eq.\ \ref{eq:24}), which  thus grows unboundedly in the thermodynamic 
limit. This is of course radically  different from the one-body  case in any finite dimension $d$
(although the limit of infinite coordination number is familiar in the different context of 
dynamical mean-field theory~\cite{dmftgeorgeskotliar} in $d=\infty$). A theory of localization in Fock-space must 
thus be able to explain how the occurrence of a divergent coordination number is in effect mitigated, such that an 
MBL transition exists in the thermodynamic limit $N \rightarrow \infty$. We will suggest one way to do so in subsequent 
work.~\cite{SWDEL2}


\begin{acknowledgments}
Many helpful discussions with John Chalker, Sthitadhi Roy and Peter Wolynes are gratefully acknowledged. 
We also thank the EPSRC for support under grant EP/L015722/1 for the TMCS Centre for Doctoral Training, and 
grant EP/N01930X/1.
\end{acknowledgments}


\bibliography{paper}

\begin{thebibliography}{16}
\bibitem [{FN4()}]{FN4}%
  \BibitemOpen
  \href@noop {} {}\bibinfo {note} {For 1BL the eigenvalue spectrum is not of
  course Gaussian, but the considerations here apply to the first and second
  moments of it, which is sufficient for our purposes.}\BibitemShut {Stop}%
\bibitem [{FN5()}]{FN5}%
  \BibitemOpen
  \href@noop {} {}\bibinfo {note} {It is in fact well converged to the form
  shown after many fewer disorder realisations than the $200$ used to generate
  fig.\ \ref{fig:fig3}. As expected from self-averaging, each realisation also
  yields essentially the same eigenvalue spectrum, modulo mild differences
  attributable to the modest system size.}\BibitemShut {Stop}%
\end{thebibliography}

\end{document}